\documentclass{article}

\usepackage{amsmath}
\usepackage{amssymb}
\usepackage{float}
\usepackage[T1]{fontenc}
\usepackage{graphicx}
\usepackage{subcaption}
\usepackage{authblk} 
\usepackage[top=3cm,bottom=3cm,left=3.5cm,right=3.5cm]{geometry}
\usepackage[backend=biber, style=numeric-comp,sorting=none]{biblatex}
\usepackage{hyperref}

\usepackage{makecell}
\usepackage[table]{xcolor} 
\usepackage{colortbl}
\usepackage{booktabs}
\usepackage{multirow}
\usepackage[utf8]{inputenc}
\DeclareUnicodeCharacter{0302}{\^{}}

\title{Two Dynamical Scenarios for Binned Master Sample Interpretation}

\author[1,2]{Giovanni Montani\thanks{\href{mailto:giovanni.montani@enea.it}{giovanni.montani@enea.it}}}

\author[2,3,4]{Elisa Fazzari\thanks{\href{mailto:elisa.fazzari@uniroma1.it}{elisa.fazzari@uniroma1.it}}}

\author[1]{Nakia Carlevaro\thanks{\href{mailto:nakia.carlevaro@enea.it}{nakia.carlevaro@enea.it}}}

\author[5,6,7]{Maria Giovanna Dainotti\thanks{\href{mailto:maria.dainotti@nao.ac.jp}{maria.dainotti@nao.ac.jp}}}

\affil[2]{\itshape Physics Department, Sapienza University of Rome, P.le A. Moro 5, 00185 Roma, Italy}
\affil[3]{\itshape Istituto Nazionale di Fisica Nucleare (INFN), Sezione di Roma, P.le A. Moro 5, I-00185, Roma, Italy}
\affil[4]{\itshape Physics Department, Tor Vergata University of Rome, Via della Ricerca Scientifica 1, 00133 Roma, Italy}
\affil[1]{\itshape ENEA, Nuclear Department, C.R. Frascati, Via E. Fermi 45, 00044 Frascati, Italy}
\affil[5]{\itshape Division of Science, National Astronomical Observatory of Japan, 2-21-1 Osawa, Mitaka 181-8588, Tokyo, Japan}
\affil[6]{\itshape Astronomy Department, The Graduate University for Advanced Studies (SOKENDAI), Shonankokusaimura, Hayama, Miura 240-0115, Kanagawa, Japan}
\affil[7]{\itshape Space Science Institute, 4765 Walnut St Ste B, Boulder, CO 80301, USA}

\begin{document}
\maketitle

\begin{abstract}
We analyze two different scenarios for the late Universe dynamics, resulting into Hubble parameters deviating from the $\Lambda$CDM, mainly for the presence of an additional free parameter, which is the dark energy parameter. The first model consists of a pure evolutionary dark energy paradigm, as result of its creation by the gravitational field of the expanding Universe. The second model also considers an interaction of the evolutionary dark energy with the matter component, postulated via the conservation of the sum of their ideal energy-momentum tensors. These two models are then compared \textit{via} the diagnostic tool of the effective running Hubble constant, with the binned data of the so-called ``Master sample'' for the Type Ia Supernovae. The comparison procedures, based on a standard MCMC analysis, led to a clear preference of data for the dark energy - matter interaction model, which is associated to a phantom matter equation of state parameter (very close to $-1$) when, being left free by data (it has a flat posterior), it is fixed in order to reproduce the decreasing power-law behavior of the effective running Hubble constant, already discussed in literature. 
\end{abstract}

\maketitle

\section{Introduction \label{sect:sec1-entropy-3796086}}
 {The} 
emergence of a $4\sigma$ discrepancy between the measurement of the Hubble \mbox{constant \cite{2000Natur.404..955D,2013ApJS..208...20B}} by the SH0ES and Planck collaborations, {refs.} 
 \cite{SH0ES, Planck2018} referred to as Hubble tension, led the community to reconsider the $\Lambda$CDM model as the only viable cosmological dynamics for the late Universe evolution \cite{whitepaper_cosmoverse, divalentino-Hubbletension, vagnozzi-seven, montani_hubbletension_1, montani_hubbletension_4, montani_hubbletension_6, montani_deangelis, schiavone2023, montani_hubbletension_9, Nojiri-odintsov, nunes2, teixeira_ddr, giare_inflation,2024ApJ...963L..43A,Scolnic_2022,2018ApJ...859..101S,Brout:2022vxf,SH0ES,Scolnic:2023mrv,Jones:2022mvo,Anand:2021sum,Freedman:2021ahq,Uddin:2023iob,Huang:2023frr,Li:2024yoe,Kourkchi:2020iyz,Schombert:2020pxm,Murakami:2023xuy,Breuval:2024lsv,SPT-3G:2022hvq,ACT:2020gnv,2024A&A...686A.210F,2024arXiv241218493L,2024arXiv241214750S}. This tension is further emphasized by several other independent and model-independent determinations of $H_0$, distinct from SNe Ia calibrated with Cepheids (see \cite{Verde_2019} for a review). These include measurements based on SNe Ia calibrated with the tip of the red giant branch (TRGB) \cite{Freedman_TRGB_1, Freedman_TRGB, Soltis_TRGB}, galaxy distances using the Tully--Fisher relation (TFR) \cite{watkins_tully, haridasu_tully}, Active Galactic Nuclei (AGN) \cite{Lu_AGN}, Gravitational Waves (GWs) and Dark Sirens \cite{gerardi_GW, Gray_GW, Li_GW, Mozzon_GW, Palmese_GW}, strong lensing time delays \cite{Zhu_lensing, Liu_lensing}, Type II \mbox{Supernovae \cite{jaeger_snii}}, Megamasers \cite{Pesce_masers}, and Surface Brightness Fluctuations \cite{Blakeslee_SBF}.

A fundamental step regarding the reliability of the $\Lambda$CDM model has been achieved by the DESI collaboration \cite{desi, desi2}, which demonstrated how the redshift profile of their Baryonic Acoustic Oscillation (BAO) data is better fitted by a Chavellier--Polarski--Linder (CPL) model \cite{CPL1, CPL2}, also dubbed the $w_0w_a$CDM scenario. This result {{de facto}}  
stated convincing evidence for an evolutionary dark energy component across the late universe (see also \cite{desi_extended, giare_overviewDDE, giare_dynamical, Giare_Robust_DDE, colgain_Dainotti,2011IJTP...50..939R,2000MPLA...15.2305D,1984MNRAS.211..277D}). However, the DESI collaboration, while partially solving the tension between the BAO and Cosmic Microwave Background (CMB) data, left entirely open the question of calibration based on the acoustic sound horizon and that of SH0ES, using Cepheids as standard candles \cite{SH0ES}. The resulting picture, emerging from the last ten years of cosmological studies, is rather confused, and it suggests that new physical effects must be included in the late universe dynamics in order for it to take a consistent shape when combining different sources belonging to different redshift regions. 

In \cite{dainotti2021, dainotti2022}, see also \cite{dainotti2025, fazzari_H0z}, these considerations led to investigating whether, in the same Type Ia Supernovae (SNe Ia) redshift distribution, an effective dependence of the Hubble constant on the considered binned representation can arise. In fact, a phenomenological power-law decreasing behavior has been detected as a better fit to binned data with respect to the $\Lambda$CDM model. 
The power-law behavior seen in the $H_0$ could be due either to the dynamical evolving dark energy model or to the $f(R)$ theory of gravity (see \cite{schiavone2023} and \mbox{also \cite{schiavone_nuovocimento}}), but it could also account for selection biases, which are hidden due to different statistical assumptions (see \cite{Dainotti2024_statassu}) and could be due to the evolution of the parameters of the SNe Ia (see \cite{Nicolas_2021}), the tip of the red giant brach stars \cite{Freedman}, and of GRBs \cite{Dainotti_2023}.

\section{Theoretical Formulation\label{sect:sec1-entropy-3796086}}
Here, we perform a study based on two different models, which are then compared with the binned data of the so-called ``Master sample'' \cite{dainotti2025}, in analogy to the study in \cite{fazzari_H0z}. The first scenario is based on an evolutionary dark energy formulation, which evolves with the redshift because it is created by the background gravitational field of the expanding universe \cite{elizalde_odintsov, fazzari_H0z, fazzari_deleo, nunes1, nunes2} (for a detailed discussion see [Schiavone et al., ``Revisiting the Matter-Creation Process: Constraints from Late-Time Acceleration and the Hubble Tension'', in preparation]), 
according to a phenomenological {{ansatz}}  
for the produced particle rate, slightly modified with respect to the studies in \cite{matcre_calvaoLima,matcre_montani2001} (actually we propose a mixed version of the two rates discussed in these papers, respectively). The second model is a revised version of the previous one, but with a completely different physical picture. In fact, the dark energy is still created by the gravitational field, but now it also interacts with the universe matter component via the condition that the sum of their energy--momentum tensors is conserved, instead of treating them separately (for other approaches with deformed matter contribution with respect to the $\Lambda$CDM model, see \cite{dainotti2024binquasars}). After testing these two models with the binned Master sample, we compare their predictions, especially in comparison to the $\Lambda$CDM dynamics and the power-law scaling, respectively.

The general scenario for developing the late universe models is a flat isotropic \mbox{picture \cite{efstathiou_planck, weinberg-grav-cosm}}, whose line element reads as 
\begin{equation}
	ds^2 = -dt^2 + a^2(t)dl^2
	\; , 
	\label{ed21}
\end{equation}
where $t$ denotes the synchronous time, $dl^2$ is the Euclidean infinitesimal distance, and $a(t)$ is the cosmic scale factor, regulating the expansion of the universe.  The cosmological dynamics is driven by a (cold dark $\rho_{dm}$ and baryonic $\rho_b$) matter energy density \mbox{$\rho_m=\rho_{dm}+\rho_b$} and a dark energy contribution, with energy density $\rho_{de}$ (we neglect here the radiation energy density, to be restored when investigating the dynamics up to the recombination era). Hence, the Friedmann equation is stated as follows:

\begin{equation}
	H^2(t) \equiv \Big(\frac{\dot{a}}{a}\Big)^2 = \frac{\chi}{3}\left( \rho_m(t) + \rho_{de}(t)\right)
	\; , 
	\label{ed22}
\end{equation}
where the dot refers to differentiation with respect to $t$ and $\chi$ denotes the Einstein constant. 

In the first proposed model, we retain the matter contribution $\rho_m$ in its standard form, i.e., governed by the dynamics $\dot{\rho}_m(t) + 3H\rho_m(t)=0$, which provides
\begin{equation}
\rho_m(z) = \rho_{m0} (1+z)^3\; , 	\label{ed23}
\end{equation}
where, here and in the whole paper, we denote with the subscript $0$ the present-day value of a quantity, and we have introduced the redshift variable $z\equiv 1/a - 1$ (we set equal to unity the present-day scale factor value). As a modification with respect to the standard $\Lambda$CDM model, we consider a process of dark energy creation by the (time-varying) gravitational field of the expanding universe, which is assumed to take place at equilibrium. Under these hypotheses, the continuity equation for the dark energy density reads as follows:
\begin{equation}
	\dot{\rho}_{de}(t) = 
	- 3H(t) (1+w_{de})\left( 1 - \frac{\Gamma (H, \rho_{de})}{3H(t)}\right) \rho_{de}(t)
	\; , 
	\label{ed24}
\end{equation}
where $\Gamma$ is the particle creation rate \cite{matcre_calvaoLima, matcre_montani2001,matcre_nunes2015}. 
Above, $w_{de}$ is the dark energy parameter and it is a free parameter of the model, subject to the constraint $w_{de}< - 1/3$. In general, the phenomenological function $\Gamma (H,\rho_{de})$ is taken as a power-law of its own arguments \cite{matcre_calvaoLima, elizalde_odintsov}. In this respect, here, we consider the following \emph{ansatz}:
\begin{equation}
\Gamma(H,\rho_{de})= 
	\Gamma^*H\rho_{de}^{-\alpha}
	\;, 
	\label{ed27}
\end{equation}
where $\Gamma^*$ and $\alpha$ are positive constants. We consider in the rate expression a linear term in $H$, stating, at highest order, the role of time-varying gravity in creating particles. We also add a dependence of $\Gamma$ on a negative power of the dark energy density to suppress particle creation when its energy density increases. 

{Introducing} 
 the normalization $\Omega_{de}\equiv \chi \rho_{de}/3H_0^2$, where $H_0$ denotes the Hubble constant, Equation~\eqref{ed24} can be rewritten as 
\begin{equation}
	\Omega_{de}^{\prime}(z)= 
	3(1+w_{de})\left( 
	1 - \bar{\Gamma}\,\Omega_{de}^{-\alpha}\right)\Omega_{de}/(1+z)
	\; , 
	\label{ed28}
\end{equation}
where the prime indicates differentiation with respect to the redshift $z$ and we have defined $\bar{\Gamma}=(\Gamma^*/3)(\chi /3H_0^2)^{\alpha}$. Furthermore, accordingly defining $\Omega_{m0}= \chi \rho_{m0}/3H_0^2$, the Friedmann equation Equation (\ref{ed22}), via Equation (\ref{ed23}), stands as follows:
\begin{equation}
	E^2 (z) \equiv 
	\Big(\frac{H}{H_0}\Big)^2 = \Omega_{m0}(1+z)^{3} + \Omega_{de}(z)\; ,
	\label{ed26}
\end{equation}
with $E(z)$ denoting the universe expansion rate. Thus we get the following initial condition $\Omega_{de}(0) = 1-\Omega_{m0}$, and Equation (\ref{ed28}) now admits the following solution:
\begin{equation}
	\Omega_{de}(z) = \left[ 
	\bar{\Gamma} +  \left( 
	(1-\Omega_{m0})^{\alpha} - \bar{\Gamma}\right)\,(1+z)^{3(1+w_{de})\alpha}\right]^{1/\alpha}\;. 
	\label{ed39}
\end{equation}
We finally get, associated with our first model, the following Hubble parameter:
\begin{equation}
H(z) = H_0\sqrt{\Omega_{m0}(1+z)^3 + \left[ 
\bar{\Gamma} + \left( 
(1-\Omega_{m0})^{\alpha} - \bar{\Gamma}\right)(1+z)^{3(1+w_{de})\alpha}\right]^{1/\alpha}}\; .
	\label{ed310}
\end{equation}
This evolutionary dark energy model generalizes the $\Lambda$CDM dynamics, and it contains five free parameters, i.e., $H_0$, $\Omega_{m0}$, $\bar{\Gamma}$, $\alpha$, and $w_{de}$. It is immediately recognizable that the Hubble parameter in Equation (\ref{ed310}) reduces to the standard $\Lambda$CDM one when we require $w_{de}=-1$. 
\\
{Finally}, it is easy to realize that for $z>1$ the $H(z)$ is very weakly sensitive to the values taken by the parameter $\alpha$, and, in what follows, we will address the simplest case $\alpha = 1$. It is worth noting that, in this case, the effective equation of state for the dark energy is associated, in $z=0$, to the following parameter $w_0^{e\!f\!f}$:
\begin{equation}
w_0^{e\!f\!f} = -1 + (1+w_{de})\Big(1-\frac{\bar{\Gamma}}{\Omega_{de}(0)}\Big)
\, .\label{weff} 
\end{equation} 
Now we require that, today, the intrinsic nature of dark energy, traced by $w_{de}$, remains the same, quintessence or phantom, also for the effective equation of state, dictated by $w_0^{e\!f\!f}$. To this end, we have to require the constraint $\bar{\Gamma}\Omega _{de}(0)<1$. Since the value of $\Omega_{de}(0)$ is expected (see also the concordance hypothesis \cite{Planck2018}) close to the value $0.7$, in what follows, we will consider for $\bar{\Gamma}$ the reference value $0.5$. 

Let us now introduce the diagnostic tool, corresponding to the effective running Hubble constant with the redshift $\mathcal{H}_0(z)$ (see \cite{dainotti2021, dainotti2022, dainotti2025, fazzari_H0z, schiavone_nuovocimento}), i.e.,
\begin{equation}
	\mathcal{H}_0(z) \equiv 
	\frac{H(z)}{\sqrt{\Omega_m^0(1+z)^3 + 1-\Omega_m^0}} 
	\, , 
	\label{ed313}
\end{equation}
which, in the analyzed case $\alpha=1$, takes the following form:
\begin{equation}
	\mathcal{H}_0(z)\equiv H_0 
	\sqrt{\frac{\Omega_m^0(1+z)^3 + \bar{\Gamma} + (1-\Omega_m^0-\bar{\Gamma})(1+z)^{3(1+w_{de})}}{\Omega_m^0(1+z)^3 + 1-\Omega_m^0}}
	\, .
	\label{ed314}
\end{equation}

Before moving to the data analysis, let us now introduce an alternative version of dynamics, our second model, in which the matter and the dark energy interact and the latter is still created by the gravitational field of the expanding universe (for other interaction mechanisms and related reviews, see \cite{2000PhRvD..62d3511A,Bolotin_632,Wang_633,Wang_563,Das_564,Pan_567,Barrow_568,Amendola_569,He_570,Valiviita_571,Yang_573,Majerotto_574,Valiviita_575,Chimento_576,He_578,Chimento_579,Costa_585,Yang_588,Nunes_592,D_Amico_605,Paliathanasis_612,Cheng_621,Zhai_2023,Johnson_630,Kumar_631,Yang_615,von_Marttens_610}). The interaction between these two basic components is phenomenologically described by requiring that the sum of the matter ideal energy--momentum tensor $T_{\mu\nu}^{(m)}$ and that of dark energy $T_{\mu\nu}^{(de)}$ is conserved instead of the separated standard laws; i.e., we require that the following relation holds:
\begin{equation}
	\nabla_{\nu}\left( 
	T_{\mu }^{(m)\nu} + 
	T_{\mu}^{(de)\nu}\right) = 0
	\;.
	\label{ed315i}
\end{equation}
Defining $\Omega\equiv \Omega _m + \Omega_{de}$, in the case of the considered flat isotropic universe, the relation above can be restated via the following continuity equation: 
\begin{equation}
\Omega ^{\prime}(z)= (3\Omega(z)+3w_{de}\Omega_{de}(z))/(1+z)\;,\label{ed316i}
\end{equation}
where by construction $\Omega(0)=1$, while retaining the same process of dark energy present in Equation (\ref{ed28}), now takes the simplified form
\begin{equation}
\Omega_{de}^{\prime}(z)=-3(1+w_{de})\bar{\Gamma}\Omega_{de}(z)^{1-\alpha}/(1+z)\;,
\label{ed16bi}
\end{equation}
with the same relation $\Omega_{de}(0)=1-\Omega_{m0}$. This equation admits now the solution
\begin{equation}
	\Omega_{de}(z)= 
	\left[ (1-\Omega_{m0})^{\alpha} - 
	3\alpha(1+w_{de})\bar{\Gamma}\ln{(1+z)}\right]^{1/\alpha}
	\,,
	\label{solx}
\end{equation}
providing an alternative form of the dark energy component evolution. Clearly, the first equation using \eqref{ed16bi} can be explicitly rewritten for the matter component as
\begin{equation}
\Omega_{m}^{\prime}(z)=3\Omega_{m}(1+z)^{-1}+3(1+w_{de})\Omega_{de}(1+z)^{-1} (1+\bar{\Gamma}\Omega_{de}^{-\alpha})\;.
\label{ed16ci}
\end{equation}
Finally, we can write the following alternative effective Hubble constant:
\begin{equation}
\tilde{\mathcal{H}}_0(z)=H_0
\sqrt{\frac{\Omega}{\Omega_m^0(1+z)^3 + 1-\Omega_m^0}}
\;, 
\label{edih}
\end{equation}
associated with the equation for $\Omega$ in correspondence with the expression for $\Omega_{de}$. Also in this case, we deal with the simplest case $\alpha=1$. Despite the fact that in the present scenario the value $\bar{\Gamma} = 0.5$ is no longer essential to ensure that the effective equation of state parameter today has the same signature as the intrinsic one  $w_{de}$, we still preserve this reference choice to better compare the two physical cases, without and with dark energy--matter \mbox{interaction, respectively}. 
\section{Data Analysis \label{sect:sec1-entropy-3796086}}
We now perform a statistical analysis of these two models, labeled as DE and DE-DM, respectively, in order to find the free parameters of the models ($H_0$, $\Omega_{m_0}$, and $w_{de}$) that optimize the probability of finding the data we use. In particular, we compare the theoretical expressions of the effective running Hubble constant in \mbox{{Equations} 
 \eqref{ed314} and \eqref{edih}} with data from the \textit{{Master binned SNe Ia sample}}  
 \cite{dainotti2025}, which is a compilation of SNe Ia from DESy5 \cite{DESy5}, JLA \cite{JLA}, Pantheon+ \cite{Scolnic_2022, Brout:2022vxf}, and Pantheon \cite{2018ApJ...859..101S} without duplicates. We use this dataset, selecting the 20 equally populated bin combination, as performed in~\cite{fazzari_H0z}. 
The entire catalog covers redshifts from $0.00122$ to $2.3$. The mean redshift of the first bin is $0.0091$, while that of the last bin is $1.54$. This sample was originally fitted with a power-law (PL) function of the form $\mathcal{H}_0(z)=\frac{H_0}{(1+z)^a}$, yielding best-fit parameters $a = 0.010$ and $H_0 = 69.869$ km/s/Mpc. Both \cite{dainotti2025,fazzari_H0z} have shown that this phenomenological function is preferred over $\Lambda$CDM, $w$CDM, and a reduced version of the $w_0w_a$CDM model. 
This dataset is hereafter referred to as the “Master bin.” The uniform priors used in this work are $H_0=\mathcal{U}[60, 80]$, $\Omega_{m_0}=\mathcal{U}[0.01, 0.99]$, and $w_{de}=\mathcal{U}[-3, 1]$.

We perform the statistical analysis by sampling the posterior distribution using the Monte Carlo Markov Chain (MCMC) method implemented in the publicly available \texttt{Cobaya} {software (3.5.5 version)} 
 \cite{cobaya}. Convergence of the chains is determined using the Gelman–Rubin criterion, requiring $R-1 < 0.01$ \cite{gelman_rubin}. Statistical results and graphs are produced with the \texttt{GetDist} {tool (1.6.1 version)} 
 \cite{getdist}. Specifically, we use a preliminary version of a code that will be publicly released in a forthcoming work [Giarè, Fazzari, in~prep.].

To compare the models, we evaluate the differences in the Bayesian Information Criterion (BIC) \cite{wagenmakers2007} for the tested model with respect to the PL parametrization,  which we adopt as the reference. This difference is defined as $\Delta \mathrm{BIC} = \mathrm{BIC}_{i} - \mathrm{BIC}_{\mathrm{PL}}$. To interpret the strength of the evidence, we use Jeffreys’ scale \cite{Bayes_trotta, kass_bayes, jeffreys_scale}, which categorizes support against a model as inconclusive for $0 < |\Delta \mathrm{BIC}| < 1$, weak for $1 < |\Delta \mathrm{BIC}| < 2.5$, moderate for $2.5 < |\Delta \mathrm{BIC}| < 5.0$, and strong for $|\Delta \mathrm{BIC}| > 5.0$. Notably, negative values of $\Delta \mathrm{BIC}$ indicate a preference for the tested model over the PL parametrization. 

\section{Results\label{sect:sec1-entropy-3796086}}
\begin{figure}[H]
  \hspace{-0.3cm} \includegraphics[width=0.6\linewidth]{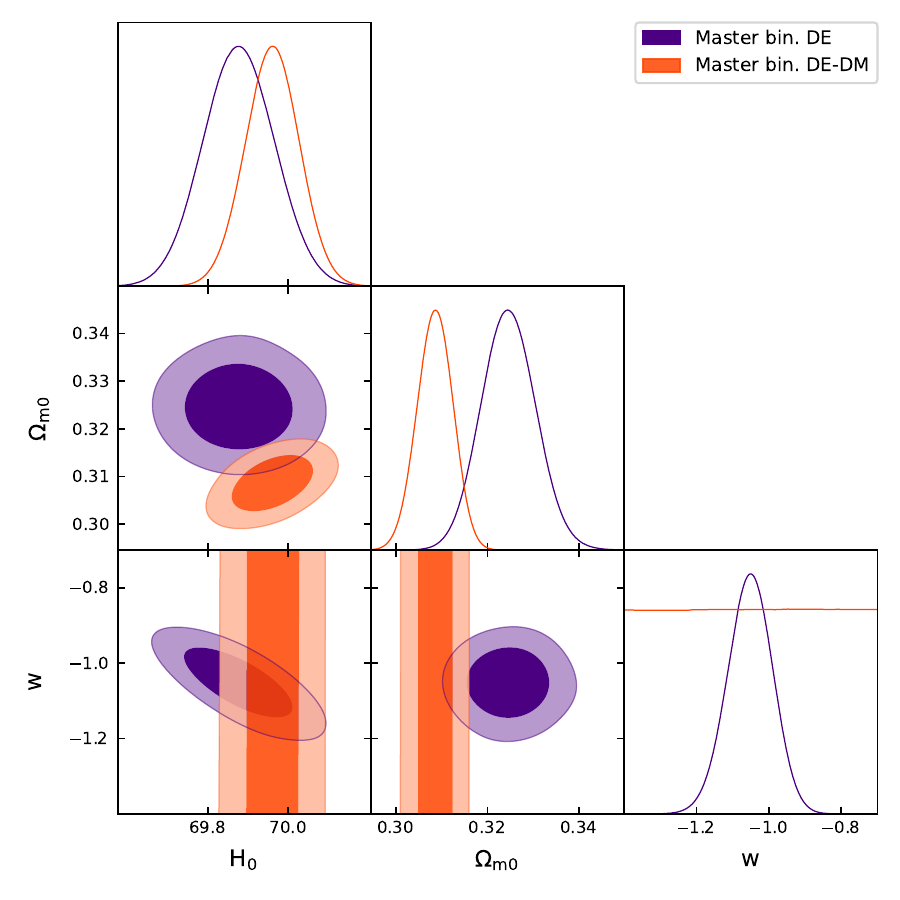}
    \caption{One-dimensional posterior probability distributions and two-dimensional $68 \%$ and $95\%$ CL contours on cosmological parameters of the DE and DE-DM models obtained using the Master binned sample.}
    \label{fig:tria_plot}
\end{figure}

{Table} 
\ref{Tab:Results_MCMC} and Figure \ref{fig:tria_plot} present the results of our analysis. These findings indicate that the Master binned data sample can effectively constrain the dark energy parameter $w_{de}$ only in the DE model, while $w_{de}$ remains unconstrained in the DE-DM model.

 {This} 
fact corresponds to the possibility of setting this parameter with a specific additional phenomenological requirement without compromising the predictive power of the data analysis.  In our case, we fix its value by imposing the significant requirement that our Hubble parameter provides a statistically meaningful representation of the power-law discussed in \cite{dainotti2025, fazzari_H0z}. 
To this end, we extract from the PL profile, reconstructed by the best-fit values recorded above, 100 points used to minimize the residuals of the two curves (resulting in residuals less than $0.15\%$). As a result of this procedure, we determine the value of $w_{de}= -1.0073$, outlining a weak phantom nature of the dark energy. 
Furthermore, for the DE model we find $\Delta \mathrm{BIC} = 6.6$, suggesting that the DE model is strongly disfavored compared to the PL case.

\begin{table}[H]
\caption{Mean values and associated uncertainties for the parameters inferred from the MCMC analysis for the DE and DE-DM models.}
\setlength\tabcolsep{0.6cm}
\label{Tab:Results_MCMC}
\centering
\begin{tabular}{lccc}
\hline\noalign{\smallskip}
\textbf{Model} & \boldmath{$H_0$} \textbf{[km s}\boldmath{$^{-1}$} \textbf{Mpc}\boldmath{$^{-1}$}\textbf{] }& \boldmath{$\Omega_{m_0}$} & \boldmath{$w_{de}$} \\
\noalign{\smallskip}\hline\noalign{\smallskip}  
DE     & $69.872 \pm 0.080$ & $0.3246 \pm 0.0053$ & $-1.052 \pm 0.058$ \\
DE-DM  & $69.959 \pm 0.064$ & $0.3084 \pm 0.0037$ & $-$ \\
\noalign{\smallskip}\hline
\end{tabular}

\end{table}

\begin{figure}[H]
  \hspace{-0.3cm}   \includegraphics[width=0.95\linewidth]{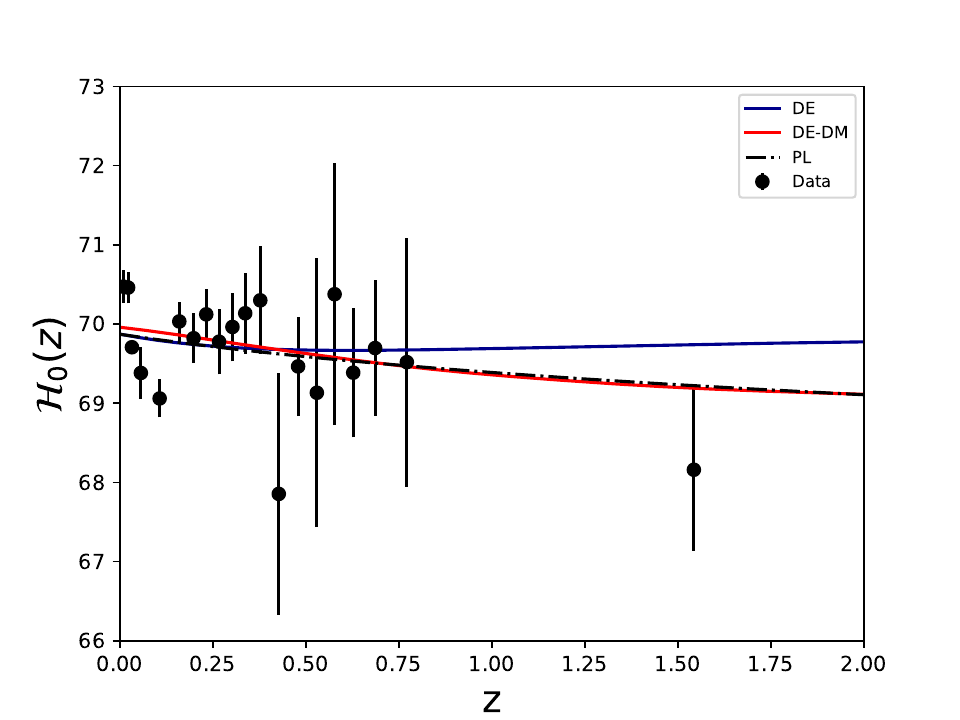}
    \caption{Reconstruction of the effective running Hubble constant for the DE model, DE-DM model, and the PL reference model. We adopted the best-fit values shown in {Equations} 
    \eqref{eq:bestfit_DE} and \eqref{eq:bestfit_DEDM}.}
    \label{fig:H0z_rec}
\end{figure}

{{Figure} \ref{fig:H0z_rec} 
shows the reconstruction of the running Hubble constant for the two theoretical models, using the following best-fit values:}\vspace{-6pt}

\begin{equation}
    \mathrm{DE:} \quad H_0= 69.872,   \quad \Omega_{m_0}=  0.3240, \quad w_{de}=-1.049 \;,
    \label{eq:bestfit_DE}
\end{equation}
\begin{equation}
    \mathrm{DE-DM:} \quad H_0= 69.959,   \quad \Omega_{m_0}= 0.3086, \, \quad w_{de}= -1.007.
    \label{eq:bestfit_DEDM}
\end{equation}
{These values are obtained by minimizing the $\chi^2$ statistic resulting from the MCMC analysis. For the DE model, we follow the standard procedure and we obtain the values shown in {Equation} 
\eqref{eq:bestfit_DE}. In the DE-DM model, since the MCMC analysis does not effectively constrain $w_{de}$, we fix it to the value obtained from the PL fit profile discussed earlier and then determine the best-fit values of the remaining parameters by performing a new MCMC analysis with $w_{de}$ held fixed.}\vspace{-6pt}

\section{Conclusions\label{sect:sec1-entropy-3796086}}
In summary, we analyzed two different reformulations of the late universe dynamics, both deviating from the $\Lambda$CDM scenario: the first model is based on pure evolutionary dark energy, in which it is created by the cosmological gravitational field; the second picture relies on dark energy--matter interaction, described via the conservation law of the sum of the two energy--momentum tensors, while dark energy now varies just by virtue of its gravitational creation. 
{The} process of dark energy creation is driven by the rate of constituent production, which is proportional to the Hubble parameter and to the inverse of the dark energy density itself: the idea we propose here is that the (non-stationary) expanding universe is able to create particles, but this process is weaker as the created energy density increases. 
{The} comparison of the two models with the $20$-bin data of the Master sample led to interesting results about how SNe Ia data alone can constrain the late universe dynamics. The comparison of the pure evolutionary dark energy model with binned data allowed us to establish that the corresponding phenomenology is not very appropriate to describe observations: actually the fit of the binned data suggested that this scenario is disfavored with respect to the power-law decreasing behavior, discussed in \cite{fazzari_H0z}. 

\indent{This} result is consistent with the idea that the evolutionary dark energy scenario is essential to properly interpret the DESI collaboration data \cite{desi, desi2}, but it has a limited impact on the behavior of the $\mathcal{H}_0(z)$, see \cite{dainotti2021, dainotti2022, dainotti2025}, which is expected to be related to the Hubble \mbox{tension itself.}   
{The} second model outlined a very different feature, since the MCMC  procedure is unable to constraint the value of the dark energy parameter $w_{de}$. This fact suggests that the value of such a parameter does not impact, in the proposed scenario, the capability of the model to fit data. Nonetheless, this degree of freedom can be used to constraint the $\mathcal{H}_0(z)$ to be a very good representation of the power-law decay observed in~\cite{dainotti2025} when the binned Master sample is concerned. 
{This} result is of particular relevance because it shows how, like the metric $f(R)$ scenario discussed in \cite{schiavone2023}, dark energy interacting with matter is a good paradigm to reproduce a monotonically decreasing behavior of the phenomenologically observed values of $H_0$. 

\indent{The} explanation of this similarity consists of the dominant role acquired by the matter term. 
If the matter term decreases with respect to a standard $\Lambda$CDM model (nonetheless the dark energy component is growing as in a quintessence scenario), soon or later, we should observe a global decreasing rescaling of $\mathcal{H}_0(z)$, which is, in this respect, very similar to the corresponding global decreasing rescaling due to the non-minimally coupled scalar field of the $f(R)$ model in the Jordan frame, as outlined in \cite{schiavone2023, montani_deangelis}.

We can conclude that the analysis of the present manuscript reinforces the idea that the interaction of the matter and dark energy components of the universe is a very promising scenario in which we could accommodate, on one hand, the evidence from the DESI collaboration observations and, on the other hand, the tension that such data generate with the SH0ES collaboration \cite{Brout:2022vxf} versus the SNe Ia calibration \cite{2018ApJ...859..101S, Scolnic_2022}, the latter effect being mainly driven by an altered matter contribution with respect to the $\Lambda$CDM paradigm.

\section*{Author Contributions}
Conceptualization, G.M.; methodology, G.M., E.F., N.C., and M.G.D.; software, E.F. and N.C.; validation, E.F. and N.C.; formal analysis, G.M., N.C., and E.F.; investigation, G.M. and N.C.; resources, M.G.D.; data curation, E.F. and M.G.D.; writing---original draft preparation, G.M., E.F., and N.C.; writing---review and editing, G.M., E.F., N.C., and M.G.D.; visualization, E.F. All authors have read and agreed to the published version of the manuscript.

\section*{Acknowledgments}
Author E. Fazzari is supported by ``Theoretical Astroparticle Physics'' (TAsP), iniziativa specifica INFN. 
E.F. acknowledges the IT Services at The University of Sheffield for providing high-performance computing resources. M.G. Dainotti acknowledges the support of the DoS and is also grateful for the support by the JSPS Grant-in-Aid for Scientific Research (KAKENHI) (A), Grant Number JP25H00675.

\printbibliography
\end{document}